\documentstyle[12pt,twoside,fleqn,espcrc1s,epsfig]{article}
\newcommand{\AmS}{{\protect\the\textfont2
  A\kern-.1667em\lower.5ex\hbox{M}\kern-.125emS}}

\title{Present Status and Future of DCC Analysis}

\author{Tapan K. Nayak\footnote{Invited talk at the Quark Matter '97,
Tsukuba, Japan, 1-5 Dec 1997}
 {\it for the WA98 Collaboration } }

\begin{document}

\maketitle

\noindent
{\small{
M.M.~Aggarwal$^{a}$, A.~Agnihotri$^{b}$, Z.~Ahammed$^{c}$,
A.L.S.~Angelis$^{d}$, V.~Antonenko$^{e}$, 
V.~Arefiev$^{f}$, V.~Astakhov$^{f}$,
V.~Avdeitchikov$^{f}$, T.C.~Awes$^{g}$, P.V.K.S.~Baba$^{h}$
% ...
}}

\smallskip
\noindent
\small\it{$^{a}$University of Panjab, Chandigarh 160014, India}
\small\it{$^{b}$University of Rajasthan, Jaipur 302004, Rajasthan, India}
\small\it{$^{c}$Variable Energy Cyclotron Centre, Calcutta 700064, India}
\small\it{$^{d}$University of Geneva, CH-1211, Geneva 4, Switzerland}
\small\it{$^{e}$RRC (Kurchatov), RU-123182 Moscow, Russia}
\small\it{$^{f}$Joint Institute for Nuclear Research, RU-141980 Dubna,
Russia}
\small\it{$^{g}$Oak Ridge National Laboratory, Oak Ridge, Tennessee
37831-6372, USA}
\small\it{$^{h}$University of Jammu, Jammu 180001, India}
\small\it{$^{i}$University of M{\"u}nster, D-48149 M{\"u}nster, Germany}
\small\it{$^{j}$SUBATECH, Ecole des Mines, Nantes, France} \\
\small\it{$^{k}$Gesellschaft f{\"u}r Schwerionenforschung (GSI), D-64220
Darmstadt, Germany}
\small\it{$^{l}$Universiteit Utrecht /NIKHEF, NL-3508 TA Utrecht, The
Netherlands}
\small\it{$^{m}$University of Lund, SE-221~00 Lund, Sweden}
\small\it{$^{n}$University of Tsukuba, Ibaraki 305, Japan}
\small\it{$^{o}$Nuclear Physics Institute, CZ-250 68 Rez, Czech Rep.}
\small\it{$^{p}$KVI, University of Groningen, NL-9747 AA Groningen, The
Netherlands}
\small\it{$^{q}$Institute for Nuclear Studies, 00-681 Warsaw, Poland}
\small\it{$^{r}$MIT Cambridge, MA 02139, USA}
\small\it{$^{s}$Institute of Physics, Bhubaneswar 751005, India}
\small\it{$^{t}$University of Tennessee, Knoxville, Tennessee 37966, USA}

\normalsize

% typeset front matter

\begin{abstract}
    Disoriented Chiral Condensates (DCC) have been predicted to form
    in high energy heavy ion collisions where the
    approximate chiral symmetry of QCD has been restored. This leads
    to large imbalances in the production of charged to neutral pions.
    Sophisticated analysis methods are being developed to
    disentangle DCC events out of the large background of events with
    conventionally produced particles. We present a short review of current 
    analysis methods and future prospects.
\end{abstract}

\section{Introduction}

The QCD phase transition from normal hadronic matter to the 
Quark-Gluon-Plasma (QGP), in case of high energy heavy ion collisions,
manifests itself in two forms:
(1) Deconfinement transition and
(2) Chiral symmetry restoration. One of the interesting
consequences of the chiral transition 
is the formation
of a chiral condensate in an extended domain, such that the direction
of the condensate is misaligned from the true vacuum direction. This
is called the formation of Disoriented Chiral Condensates (DCC). It
has been proposed that the decay of the DCC domains would lead to
large imbalances in the production of charged to neutral pions.
The task for experimentalists is to carefully measure the number of neutral
and charged pions as well as study their spectra. The challenge is to 
design sophisticated analysis tools on an
event-by-event basis to identify DCC amidst the large background
due to conventionally produced particles.

The formation of DCC domains has been proposed by Anselm\cite{ans1},
by Blaizot and Krzywicki\cite{blai1} and by Bjorken, Kowalski and
Taylor\cite{bjor1} in the context of high energy hadronic collisions
in order to explain the puzzling Centauro and anti-Centauro type of
events observed by cosmic ray experiments\cite{cosmic1}. 
Bjorken et al.\cite{bjor1,bjor2}
proposed the so called ``Baked Alaska'' model which suggests
that in case of hadronic collisions a hot fireball is produced 
with cold interiors having anomalous chiral order parameter.
Rajagopal and Wilczek\cite{raj1} were the first ones to discuss the
DCC phenomena in the context of heavy ion reactions. They have suggested
that the nonequilibrium dynamics during the chiral symmetry breaking
phase transitions in case of heavy ion collisions may produce DCC domains.
There is tremendous progress 
{\cite{bjor2,gavin,gavin2,randrup1,randrup2,asakawa,raj2}
in terms of theoretical understanding of DCC since then starting with
different formalisms.

In the framework of the linear sigma model the Lagrangian can be expressed
in terms of the order parameter $\Phi \equiv (\sigma,\vec{\pi})$, which is
a combination of the scalar field $\sigma$ and the pion field $\vec{\pi}$.
At low temperatures the chiral symmetry is spontaneously broken. The
potential, $V(\Phi)$, has a minimum in the sigma direction and all the
direction of $\vec{\pi}$ are equally populated. Thus the distribution of
neutral pion fraction defined by
\begin{equation}
f =  \frac{N_{\pi^0}}{N_{\pi^0} + N_{\pi^+}+N_{\pi^-}}
\end{equation}
is a gaussian with a mean of 1/3. This leads to the
isospin symmetry of pions. As the system goes through the chiral transition
and then rapidly expands and cools, it may roll down from the unstable
local maximum of $V(\Phi)$ to the nearly stable values of one of the pion
directions. The field will have to eventually settle in to the true ground
state, but oscillations will continue for sometime which leads to amplification
of soft pion modes. This effectively creates a cluster of low $p_T$ pions
in a correlated region (of a so called DCC domain) with the probability of
the neutral pion fraction given by:
\begin{equation}
 P(f) =  \frac{1}{2\sqrt{f}}
\end{equation}

As is evident from this discussion, pions from a DCC domain will be
emitted at low $p_T$ and will have a distinct distribution pattern
compared to the normal pion production mechanism without DCC. 

Our ability to detect
DCC domains depends on the number of domains, size of domains and number of
DCC pions emitted from the domains. If the number of domains is more than
one, then the distribution given by (2)
gets modified, and for a large number of uncorrelated
small domains, the resulting distribution becomes gaussian. The 
size of a DCC domain and the energy content determine the number of DCC pions.
Clearly the fewer the number of domains and larger the number of emitted
pions, the easier it is to detect in the laboratory. 
In addition, the probability of DCC formation in a nuclear collision
at a given energy is important for the observation of DCC.

The cleanest evidence for the DCC formation can be obtained by precision 
measurement of $N_{\pi^0}$ and $N_{\pi^\pm}$, or equivalently, 
$N_\gamma$ and $N_{\rm ch}$ distributions. This assumes all charged
particles are pions and all photons come from $\pi^0$ decays.
In this manuscript we will deal mostly with analysis of neutral and
charged particle distributions. Other signals which have been proposed
for detection of DCC will be discussed towards the end of the
manuscript.

\section{Accelerator Experiments}

The observation of centauro type of events have attracted lot of 
attention since the early 80's. 
%For the centauro events observed by Chacaltaya emulsion 
%experiment\cite{cosmic1} the number
%of charged hadrons estimated at the interaction vary between 63 and 90,
%whereas the number of gamma rays is consistent with 0. The JACEE collaboration
%has shown the classic case of an ``anti-centauro'' event where
%a region of $\eta-\phi$ phase space contains an anomalously large number
%(about 35) of photons and small number (only 1) of charged particles.
Several accelerator experiments
have been performed to search for such unusual events even
before the concept of ``DCC'' came into being.
The UA1 and UA5 experiments at the CERN-SPS have carried out search for
centauros in $p-\bar{p}$ collisions at CM energies of 540 and 900 GeV,
respectively, whereas D\O~ and CDF experiments at Fermilab have 
performed this search for CM energy of 1.8 TeV. 
These experiments have used the technique of the asymmetry of 
hadronic to electromagnetic energies. So far there is no evidence of any 
centauro type of events from these experiments.

Below we describe two current experiments for DCC search: the Minimax
experiment for $p-\bar{p}$ collisions and the WA98 experiment for heavy
ion collisions. Results from WA98 experiment will be discussed in detail.
Other experiments at CERN-SPS such as
NA45 and NA49 have plans for DCC search in near future.

\subsection{Minimax Experiment at Fermilab}

The Minimax Experiment{\cite{minimax} at Fermilab Tevatron is designed 
by Bjorken et al. for the DCC search in $p-\bar p$
collisions of 1.8TeV in the far forward region.  The experiment
is located at the C\O~ interaction region of the Tevatron and has been
designed to measure the ratio
of charged to neutral pions produced at pseudo-rapidities near 4.1.
The lego acceptance of the detector is a circle of radius 0.65 units in
$\eta$.
%The choice of the forward acceptance was motivated by the observation of
%centauro events in forward direction at large pseudorapidities.

The data analysis for DCC search in Minimax
is complicated because of small
acceptance of the detector and various efficiency factors which
come into play. The method of normalized factorial moments has been
utilized successfully to construct a set of ``robust observables'' 
which eliminate most of these effects \cite{minimax}. 
This method is described in section~5.3.
The results from
these data is consistent with no DCC production mechanism.
%These are
%the ratios of first few moments constructed from the distributions 
%of charged particles and photons. 
%These ratios yield 1 for generic distributions and 1/(i+1) for pure
%DCC. The data distributions of these robust variables are calculated
%and then compared with simulations of PYTHIA+GEANT. So far the results from
%their data is consistent with no DCC production mechanism.
%We will discuss these aspects
%briefly in section ***. More details can be found in \cite{minimax}.

\subsection{WA98 Experiment at CERN-SPS}

The WA98 experiment\cite{bolek1} at CERN-SPS emphasizes on high
precision, simultaneous measurement of both hadrons and photons.
The experimental setup consists of large acceptance hadron and photon 
spectrometers, detectors for charged particle and photon multiplicity 
measurements, and calorimeters for transverse and forward energy measurements. 
%WA98 is well suited for a detailed comparison of
%charged particle to photonic signals. 
At present our search is limited to 
detailed event-by-event analysis of photon distributions from the
Photon Multiplicity Detector and charged particles from Silicon Pad
Multiplicity Detectors.
Details of these detectors may be found elsewhere, here we
give the essential points necessary for our present discussion. \\

%Detectors in WA98 may be categorized into two types: \\
%\noindent
%(A) Detectors for electromagnetic signals are: (1) 
%      Photon Multiplicity Detector (PMD) for multiplicity of
%      photons in the range of 2.4$\leq\eta \leq$4.4, 
%   (2) a lead-glass detector array for photon and neutral hadron spectroscopy
%     (3) the mid rapidity calorimeter for transverse energy ($E_T$
%         measurement of electromagnetic particles. \\
%\noindent
%(B) Detectors for hadronic (charged) signals are:
%      (1) Silicon drift and Silicon pad multiplicity detector (SPMD) for 
%      charged particle multiplicity measurements,
%      (2) The plastic ball detector,
%      (3) A charged particle tracking system consisting of a 
%     1.6Tm dipole magnet, tracking chambers and two time-of-flight walls, and 
%      (4) Mid rapidity calorimeter for $E_T$ of charged hadrons.

%In addition several trigger detectors and a zero degree calorimeter (ZDC)
%for measuring the forward energy are employed for data cleanup and 
%event characterization. 
%WA98 is well suited for a detailed comparison of
%charged particle to photonic signals. 
%At present our search is limited to 
%detailed event-by-event analysis of charged particles from SPMD and photons
%from PMD. Details of these detectors may be found elsewhere, here we
%give the essential points necessary for our present discussion.

\noindent
{\it \bf The Photon Multiplicity Detector (PMD)}\\
The PMD consists  of a 3$X_0$ thick lead
converter sheet in front of an array of 54,000 pads, 
and is located at a distance of 21.5 meters from the target. 
Signals from several neighboring pads are combined to form clusters,
characterized by the total ADC content and the hit positions. 
%There is a
%92\% probability for photons with energy greater than 0.4 GeV to shower
%in the lead converter and produce large signal, whereas non-showering hadrons
%give 
%a signal corresponding to those of single minimum ionizing particle (MIP).
A threshold of 3 MIPs on the preshower ADC content gives an average photon 
counting efficiency in the range of 70\%-75\% depending on the centrality, 
with about 32\% contamination due to 
showering hadrons. Thus the measured particle
clusters, called ``$\gamma$-like clusters''
contain mostly photons together with a sizable contribution of charged
pions. \\

\noindent
{\it \bf The Silicon Pad Multiplicity Detector (SPMD)}\\
The SPMD is located at 32.8 cm from the target.
It is based on a double metal, AC coupled design
consisting of approximately 4000 pads, arranged radially with 180 
$\phi-$bins and 22 equal$-\eta$ bins between $\eta = 2.35$ and 3.75. 
The granularity of the detector is high enough that we have only $15-20$\%
occupancy in the highest multiplicity events. The efficiency of detecting
a charged particle in the active area has been determined in a test beam to
be 99\%. Conversely the detector is transparent to high
energy photons, since only about 0.2\% are expected to interact in the
silicon.

\section{DCC Analysis Methods} 

Most of the DCC analyses so far are based
on studying the fluctuations in neutral and charged particle
measurements. Various analysis methods are being employed to characterize
unusual events which show up beyond the statistical fluctuations.
All the analysis
in WA98 are being carried out in {\it event-by-event} basis.
Current techniques in DCC analysis are the following:
\begin{itemize}
  \item{Global event characteristics:} This analysis is performed by using
  the total number of photons and charged particles over the entire
  phase space covered by the photon and charged particle detectors.
  \item{Methods for DCC domains localized in phase space:}
    The available phase space is divided into several $\eta-\phi$ bins.    
    Here we mention two of the analysis methods which are sensitive enough to
    single out localized domains:
    \begin{itemize}
    \item{Wavelet Analysis:}  This multiresolution analysis has the capability
     to scan the entire phase space. No averaging over events or
     $\eta-\phi$ space is done.
     \item{Moments analysis:} Various moments and combinations of those are 
       calculated 
       from the distribution of photons and charged particles in each bin.
   \end{itemize}
\end{itemize}

In all these analysis, data are not corrected for detector effects
and other efficiencies. Instead we compare the data with MC simulations
which incorporates all known detector and physics effects.
The output of VENUS 4.12 event generator is passed through a full simulation
of our experimental setup using the GEANT 3.21 package from CERN.
The same method of
analysis is followed for both data and MC generated events. The idea is
to look for 
\vspace*{-0.3cm}
\begin{itemize}
  \item shapes of different distributions: non-gaussian shapes
        in case where gaussian is expected and
        differences in various distributions in case of data and MC simulation 
\vspace*{-0.4cm}\vspace{-0.4cm}
  \item exotic events: in case the shape is deviating from what is expected,
        one can study the events which show up beyond some predetermined limit.
\end{itemize}

\section{Global Event Characteristics}

The method of global event characterization in terms of the 
photon and charged particle multiplicity distributions over
the full available phase space
is suitable for the search of single large size DCC domain.
The idea here is to look for events which fall far beyond
the correlation line of these two distributions.
For the current study we have assumed that the DCC could be formed only
in central collisions. We will concentrate on the 10\% most central
events, defined by a measured transverse energy of at least 300 GeV in
$3.5< \eta < 5.5$. 

\begin{figure}[t]
\setlength{\unitlength}{1mm}
%\vspace{-0.4cm}
\begin{picture}(140,80)
\put(0,0){
\epsfxsize=7.0cm
\epsfysize=6.5cm
\epsfbox{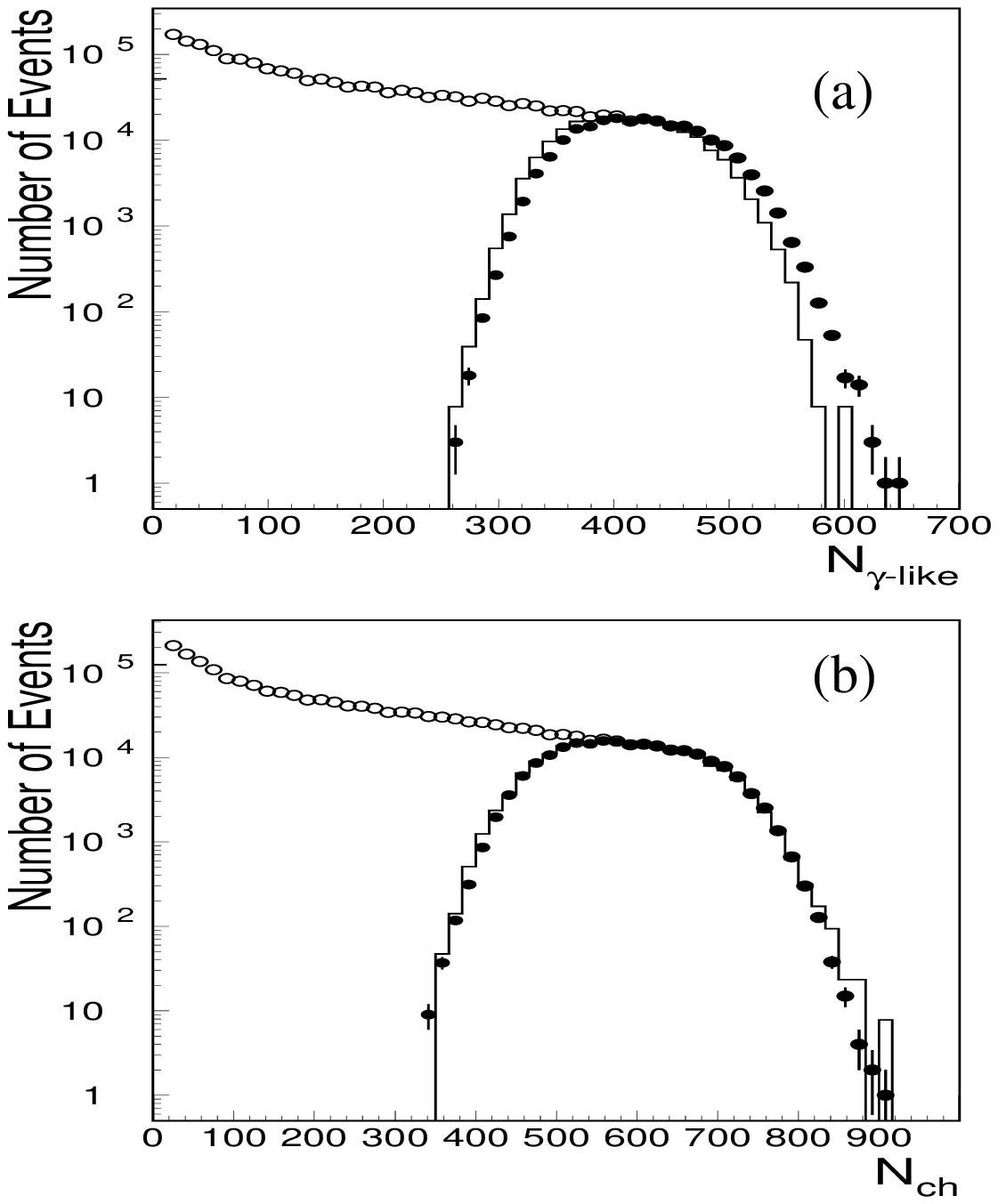}
}
\put(70,-7){
\epsfxsize=9.5cm
\epsfysize=6.7cm
\epsfbox{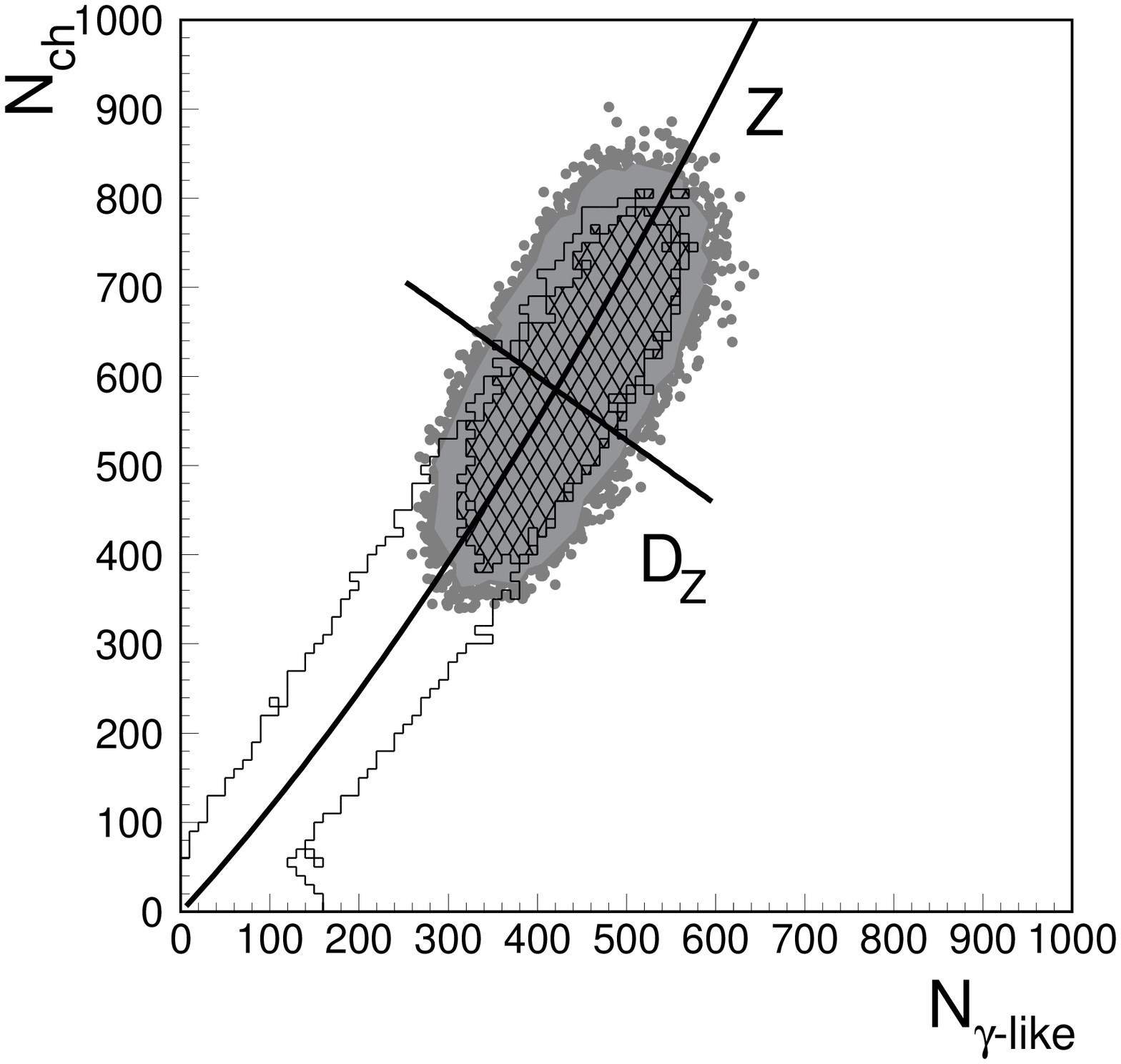}
}
\end{picture}
\vspace{-0.3cm}
\caption{
Multiplicity distributions of (a) $\gamma$-like clusters and
(b) charged particles for minimum-bias (open circles)
and central events (filled circles). The histograms show 
MC simulation results for central events.
The right hand panel shows the correlation of $N_{\gamma-{\rm{like}}}$ 
and $N_{ch}$ for central events. The hatched region is for MC simulation
results.
}
\end{figure}

\begin{figure}[t]
\setlength{\unitlength}{1mm}
\vspace{-0.7cm}
\begin{picture}(140,80)
\put(0,0){
\epsfxsize=7.5cm
\epsfysize=7.0cm
\epsfbox{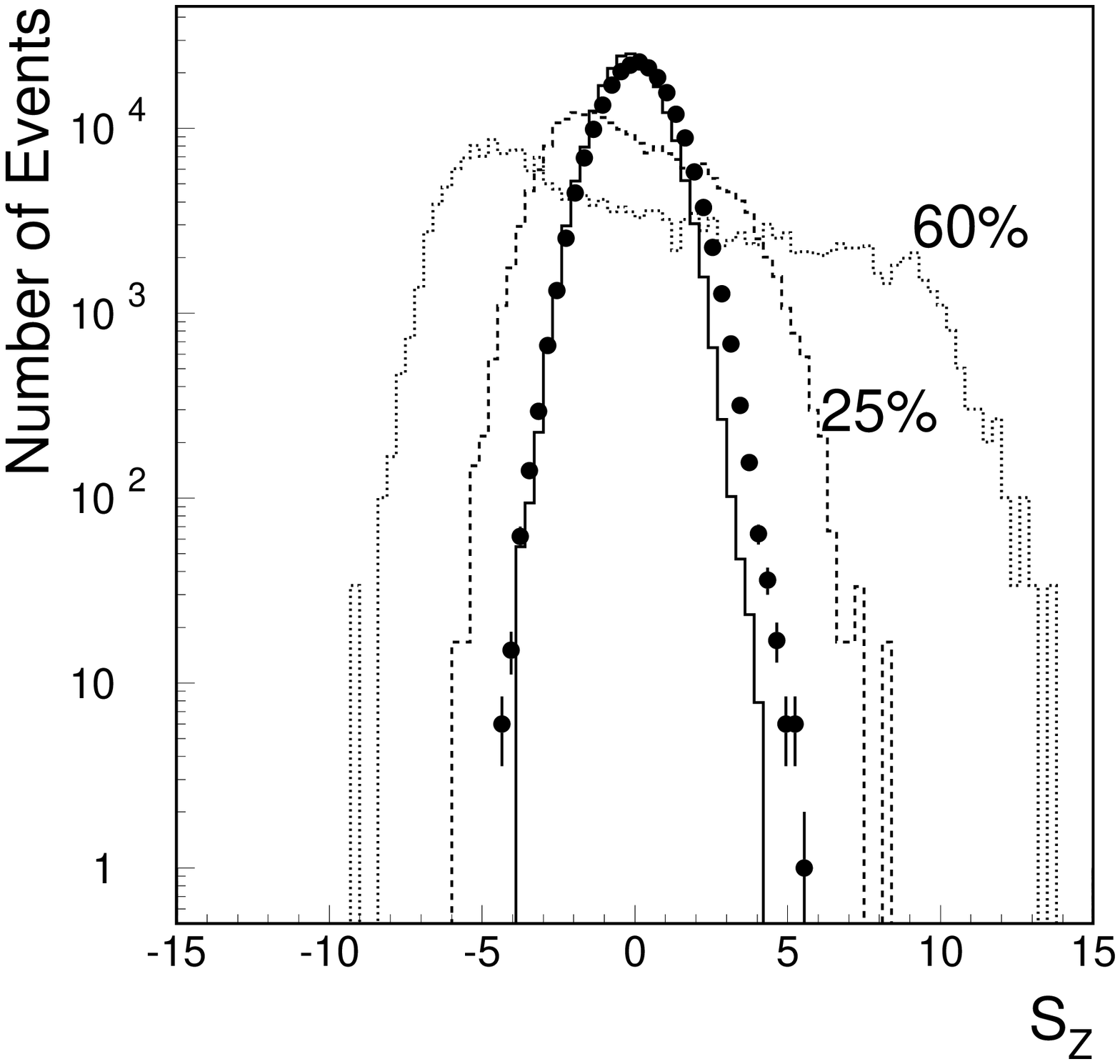}
}
\put(80,0){
\epsfxsize=7.5cm
\epsfysize=7.0cm
\epsfbox{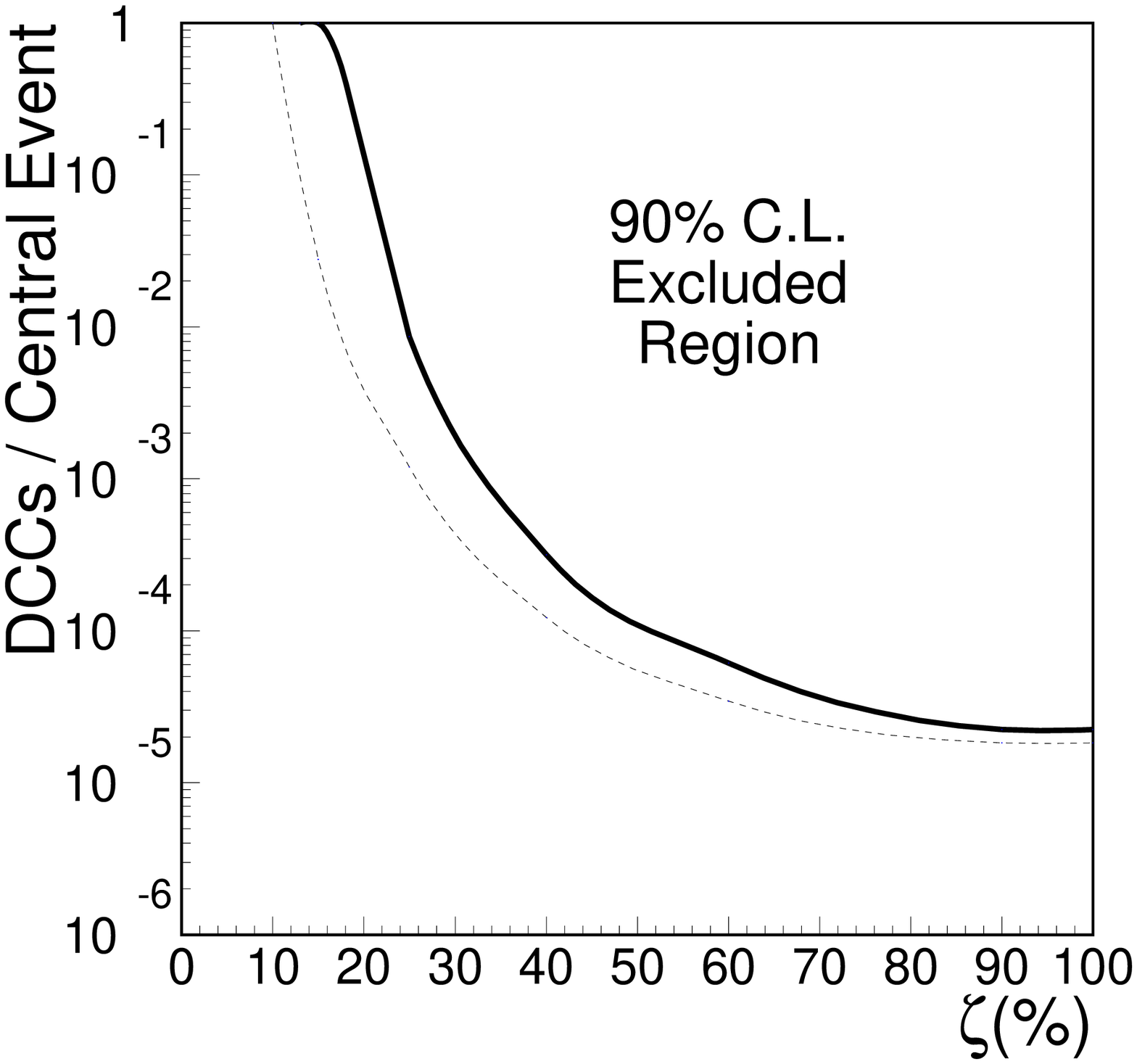}
}
\end{picture}
\vspace{-0.9cm}
\caption{
(a) $S_Z$ distribution for the experimental data is shown, 
overlaid with simulation results incorporating 
0\%, 25\% and 60\% DCC in every event.
(b) 90\% C.L
upper limit on DCC production per central event as a function of
the fraction of DCC pions under two assumptions.  
}
\end{figure}

In Figures 1a and 1b we present the minimum-bias multiplicity
distribution for $\gamma$-like clusters and charged particles.
The central data sample is shown by filled circles in these two
figures. After all cuts are applied, there are 212646 events in this sample.
A comparison with the MC simulation events chosen by identical
cuts for central events is shown by the histogram.

The correlation between the charged and neutral multiplicities is
presented on the right side of Figure~1 with the minimum bias distribution
outlined, the central MC simulated events hatched, and the central data events
shown as scattered points, each point corresponding to a single event.
A strong correlation is seen between charged and neutral multiplicities,
which suggests a more appropriate
coordinate system with one axis being the measured correlation axis
and the other perpendicular to it. This is represented by the $Z$ axis and
the $D_Z$ axis as shown in figure~1b. We have chosen to work with 
the scaled variable
$S_Z{\equiv}D_Z/\sigma_{D_Z}$ in order to compare relative fluctuations
at different multiplicities.
$S_Z$ distribution for the data is shown as filled circles in figure~2(a).

\subsection{DCC simulation}

To estimate the effect of DCC production we have modified the output of
VENUS events to include characteristic fluctuations in the relative
production of charged and neutral pions.
We assume that only a single domain of DCC is
formed in each central collision.  
A certain fraction,  $\zeta$, of the 
VENUS pions is associated with this domain
and a value of $f$ is chosen
randomly according to the distribution shown in equation 2.
Then the charges of the pions are interchanged pairwise
($\pi^{+}\pi^{-}$ or $\pi^{o}\pi^{o}$)
until the charge distribution matches the chosen value of $f$.
This simulates a DCC accompanied by the normal hadronic background
in a way that conserves energy, momentum, and charge.
The $S_Z$
distribution for the $\zeta$ = 0\%, 25\% and 60\% DCC hypotheses are
shown in figure~2. The distributions get
wider as $\zeta$ is increased. Thus DCC events would appear
as outliers with respect to the bulk of the data.

\subsection{DCC upper limit}

We expect the DCC events to show up as non-statistical tails 
in the $S_Z$ plot. Since we do not see no such events in our data sample,
we are faced with
the possibilities that single-domain DCCs are very rare, very small, or both.
To check which hypotheses are consistent with our data,  we determine 
upper limits on the frequency of DCC production as a function of
its size, as represented by $\zeta$.  

%The cut $|S_Z| > S_{\rm{cut}}$ defines a
%two-dimensional region in the scatter plot in which all events
%are considered to be ``DCC candidates''.  Once the cut is set,
%the DCC efficiency is defined by $N$(over cut)/$N$(total in simulation),
%which is a function of both the DCC fraction and the cut position.
%The background is determined by a Gaussian fit to the MC simulations
%with no DCC. With the efficiency and background determined,
%we calculate the Poisson upper limit $N_{U.L.}$ for a 90\% confidence level,
%which is $\approx$2.3 if there are no measured events 
%over the cut and no  background events are expected.
%These three numbers are combined into an upper limit, for 
%$N_{\rm{Data}}$ events, via the formula:
%\begin{equation}
%\frac{N_{\rm{DCC}}}{N_{\rm{Central}}}(S_{\rm{cut}},\zeta)
%\leq
%\frac{N_{U.L.}}{\epsilon(S_{\rm{cut}},\zeta)}
%\frac{1}{N_{\rm{Data}}}.
%\end{equation}

We have calculated the 90\% confidence limits for two scenarios and 
the results are shown in the right side of figure~2. 
The first, shown as the solid line, is based
upon the conservative assumption that the MC simulation
should describe the data perfectly in the absence of a DCC signal.
The second scenario, shown as the dashed line, assumes
that the difference between the data and VENUS is due to detector
effects and that the widths should be the same.
Further details can be found in \cite{wa98dcc}.

\section{Methods for DCC domains localized in phase space}

     The localized DCC analyses take advantage of the predictions
     that the coherent fluctuations caused by DCC tend to cluster in
     phase space. A similar analysis
     technique that described in the last section can be used
     in small phase space. Here we describe two other
     analysis methods, {\it viz.}, use of wavelets and factorial 
     moments.

\begin{figure}[t]
\vspace*{-1.0cm}
\centerline{\psfig{figure=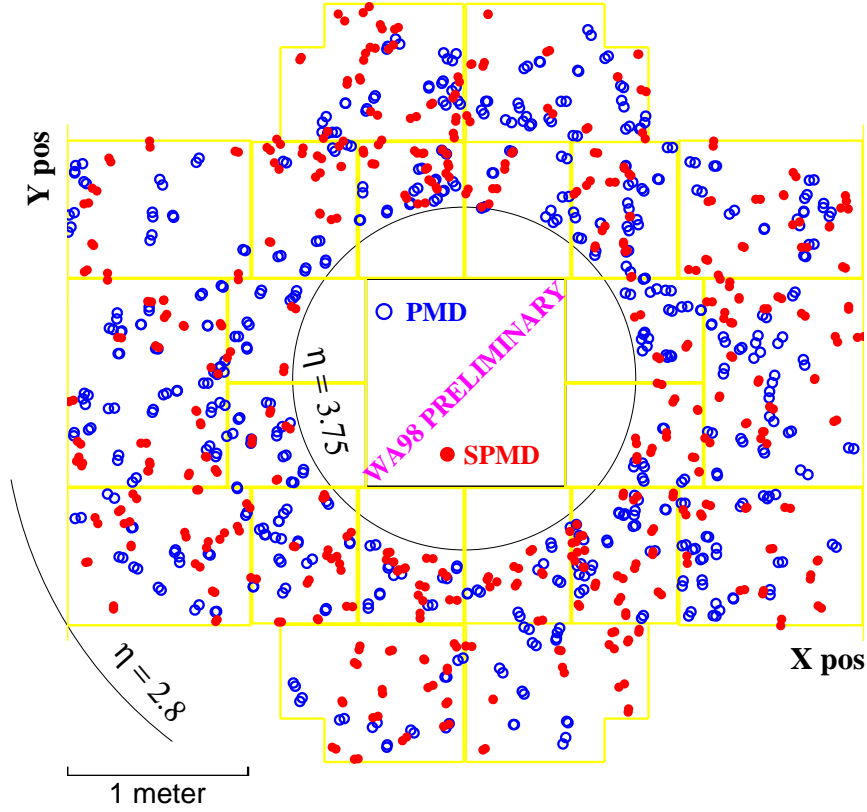,width=14cm}}
\vspace{-2.3cm}
\caption{
   $x-y$ phase space distribution of a normal event from the WA98 experiment on
   Pb+Pb reactions at 158$\cdot$A GeV/c.
}
\end{figure}

     In WA98, the localized analysis is done by taking the
     data of the full overlap regions of PMD and SPMD, which lies
     between $\eta=2.8$ and 3.75.
     In figure~3 this zone is shown for a single event in
     terms of $x-y$ phase space distribution on the PMD plane.
     Open circles show the distribution of $\gamma$-like clusters
     from PMD and filled circles show the charged particle
     distributions from SPMD projected on to PMD plane.
     The neutral and charged particle distributions are evenly
     distributed in this event. The methods discussed below are sensitive to
     single out exotic events out of a large data sample.

\subsection{DCC simulation for localized analysis}

    For the simulation of normal background events we have adopted the
    same simulation method as described in section~3 whereas 
    the DCC simulation is done differently compared the global case.
    Since the localized methods are sensitive to smaller
    phase space zones, we have the option of introducing single or
    multiple domains of varying sizes to any phase space
    into the simulation and carry out further analysis. For the present 
    study, we have chosen a DCC domain
    to be within $3\leq\eta\leq 4$ and $0\leq\phi\leq 90$.
    The common phase space of PMD and SPMD overlaps with this
    choice of domain. The rest of the analysis follows as described in section
    4.1 including the full GEANT simulation of WA98 detectors.

\subsection{Wavelet Analysis}

A unique analysis method based on the discrete wavelet transformation 
(DWT) is adopted to search for the fluctuation in neutral pion fraction 
in a localized $\eta-\phi$ space. 
This method was first proposed by 
Huang et al.\cite{huang} for DCC search. Wavelets are the basis
functions in some representations of arbitrary functions that 
satisfy certain requirements like invertibility, orthogonality 
etc. (as sines and cosines in case of Fourier transformation). These 
arbitrary functions can be functional representations of a given 
data set. The most interesting feature of the discrete wavelet transformation
is the zooming action at each location of the various resolution
scale. Due to the completeness and orthogonality of the DWT basis, 
there is no information loss at any scale. 

In our present analysis, we have chosen the arbitrary function
to be the neutral pion fraction given in equation (1). 
The DWT representation of the neutral pion fraction can be represented by,
\begin{equation}
     f^{(j)}(x) = \sum^{2^{j-1}-1}_{k=0}  f_{j-1,k}\phi_{j-1,k}(x) 
                   + \sum^{2^{j-1}-1}_{k=0}  \tilde{f}_{j-1,k}\psi_{j-1,k}(x)
\end{equation}
where $f_{j,k}$ are the mother function coefficients (MFC) and
$\tilde{f}_{j,k}$ are the father function coefficients (FFC)
at position index $k$ of resolution scale $j$. 
The MFC's and FFC's are the carriers of information at each scale. 
In Haar basis, MFC's are the average between two bins and FFC's
are the half difference between two bins. 
The variable $x$ can be $\eta$, $\phi$ or a combination of both 
$\eta$ and $\phi$. We have chosen $x$ to be $\eta$ in our case.
$\phi(\eta)$ and $\psi(\eta)$ are called the mother and father
functions, respectively. In the present analysis we have used D-4 wavelet
basis.

\begin{figure}[t]
\setlength{\unitlength}{1mm}
\vspace{-1.0cm}
\begin{picture}(140,80)
\put(0,0){
\epsfxsize=8.0cm
\epsfysize=9.0cm
\epsfbox{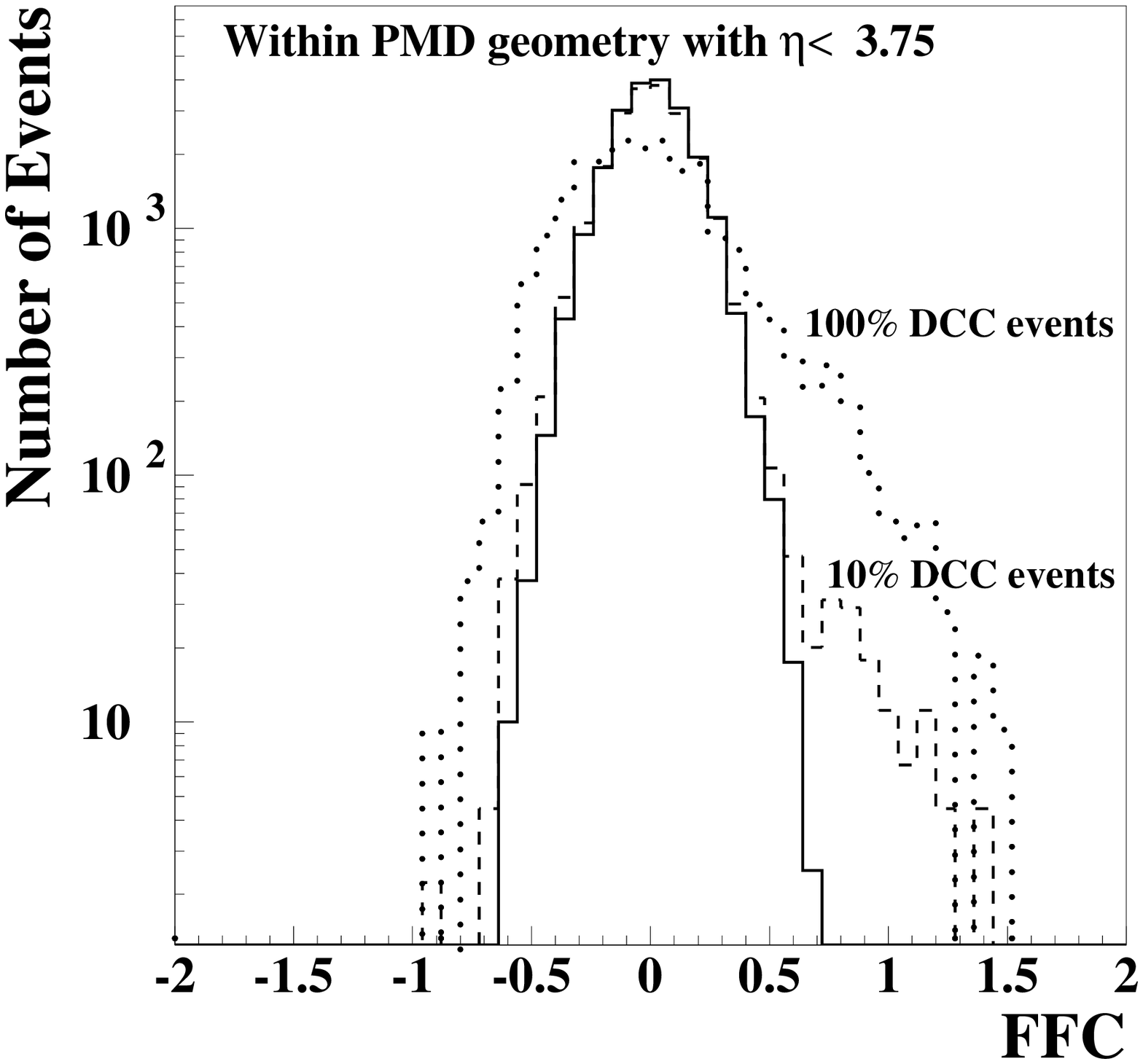}
}
\put(80,0){
\epsfxsize=8.0cm
\epsfysize=9.0cm
\epsfbox{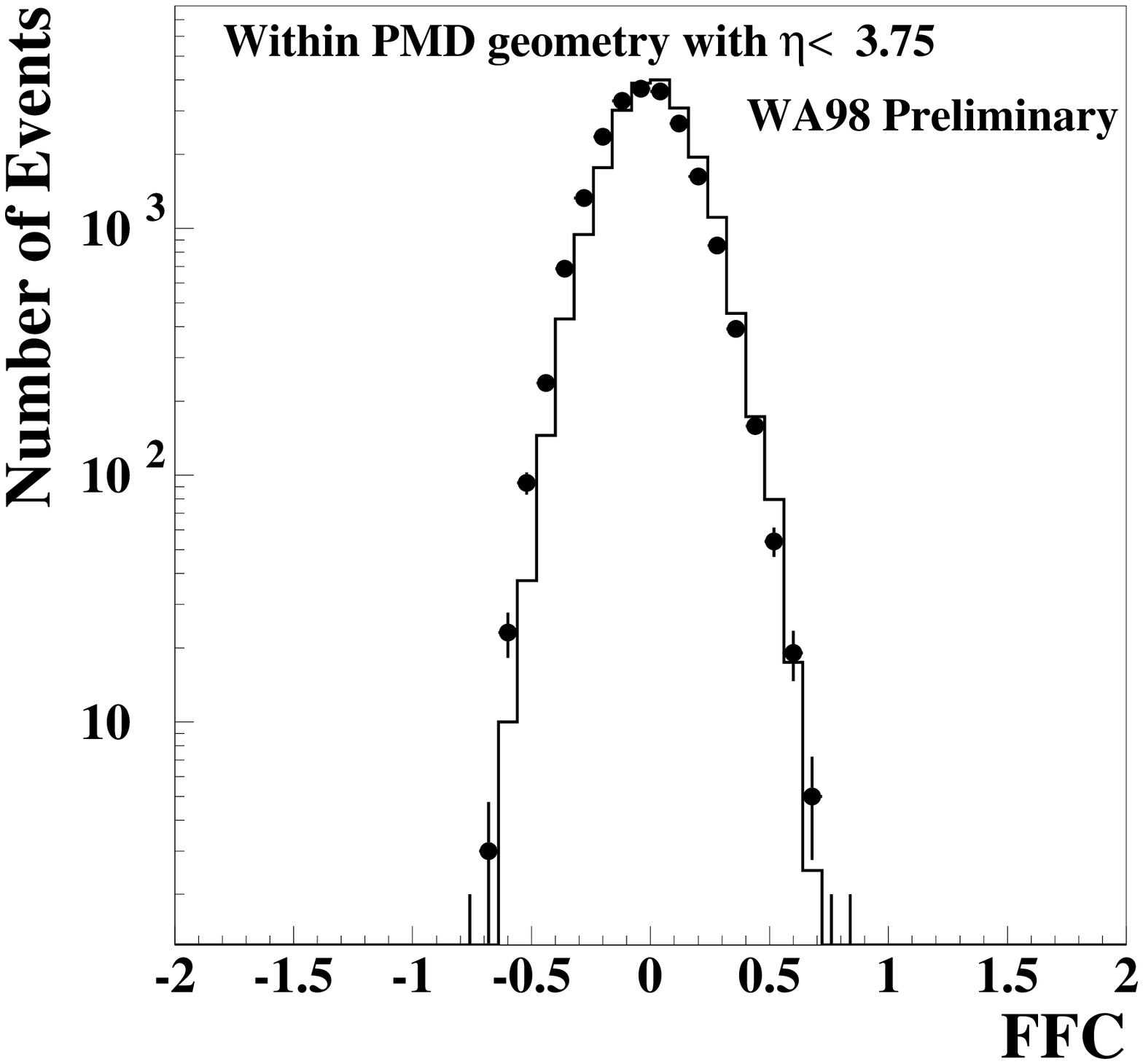}
}
\end{picture}
\vspace{-1.3cm}
\caption{
(a) FFC distribution of simulated data of different percentage 
    of DCC events.
(b) FFC distribution of data and comparison with MC simulation.
}
\end{figure}
In order to verify if the FFC distributions are sensitive to events
with DCC pions we have carried out the MC simulation described above
for different percentage of DCC events present in the full sample.
We show two of these cases in figure~4(a). The solid histogram
shows the FFC distribution at $j=2$ in the absence of any DCC event. This
is our background simulation. The dotted and dashed curves represent
the cases where DCC events are present in all the (100\%) events and
10\% of the events, respectively. In all cases the DCC domain is chosen
to be the same, as described in previous section. Significant difference
between the background histogram and events with DCC can be seen here.
In this method one is sensitive to a small admixture of DCC events
in the full sample. It is seen that even if the occurrence of DCC events 
compared to the total number of events is very less, still then there are 
significant number of events beyond $3\sigma$ of FFC distribution of 
background events at j=1. These events are to be tagged and studied in
detail for the presence of DCC.

The WA98 data FFC distribution (at $j=2$) 
for central events is shown as solid
points in figure~4(b) and overlaid on the top is the background
MC simulation. Data and simulation match quite well. 
Detailed analysis is in progress by rotating the full phase space
by a given angle in $\phi$ and repeating the whole analysis. Another way is
to rotate the SPMD plane with respect to PMD plane and determining FFC
from the resulting distribution. 

\begin{figure}[t]
\setlength{\unitlength}{1mm}
\vspace{-2.2cm}
\begin{picture}(140,90)
\put(0,0){
\epsfxsize=7.5cm
\epsfysize=8.0cm
\epsfbox{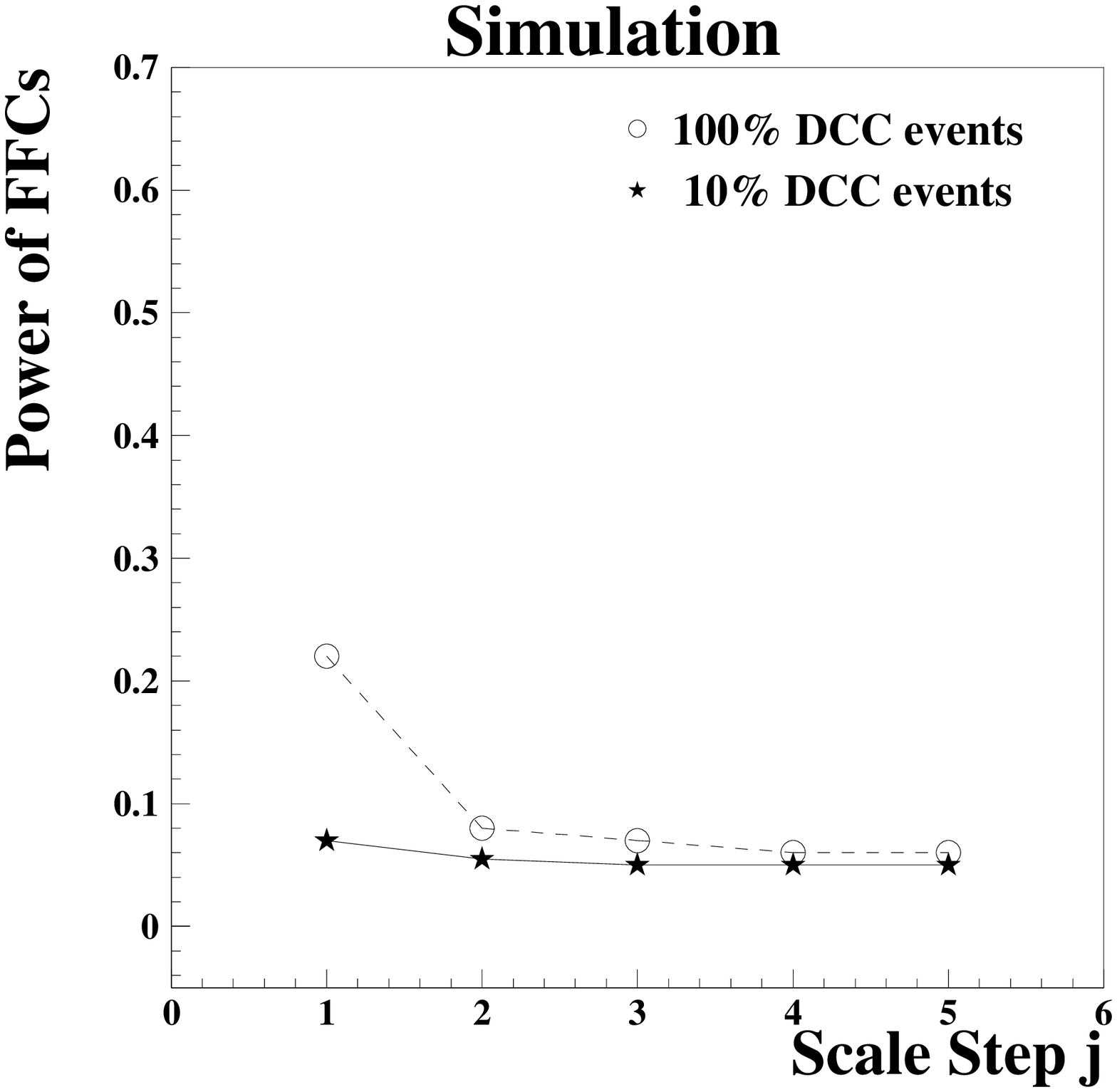}
}
\put(80,0){
\epsfxsize=7.5cm
\epsfysize=8.0cm
\epsfbox{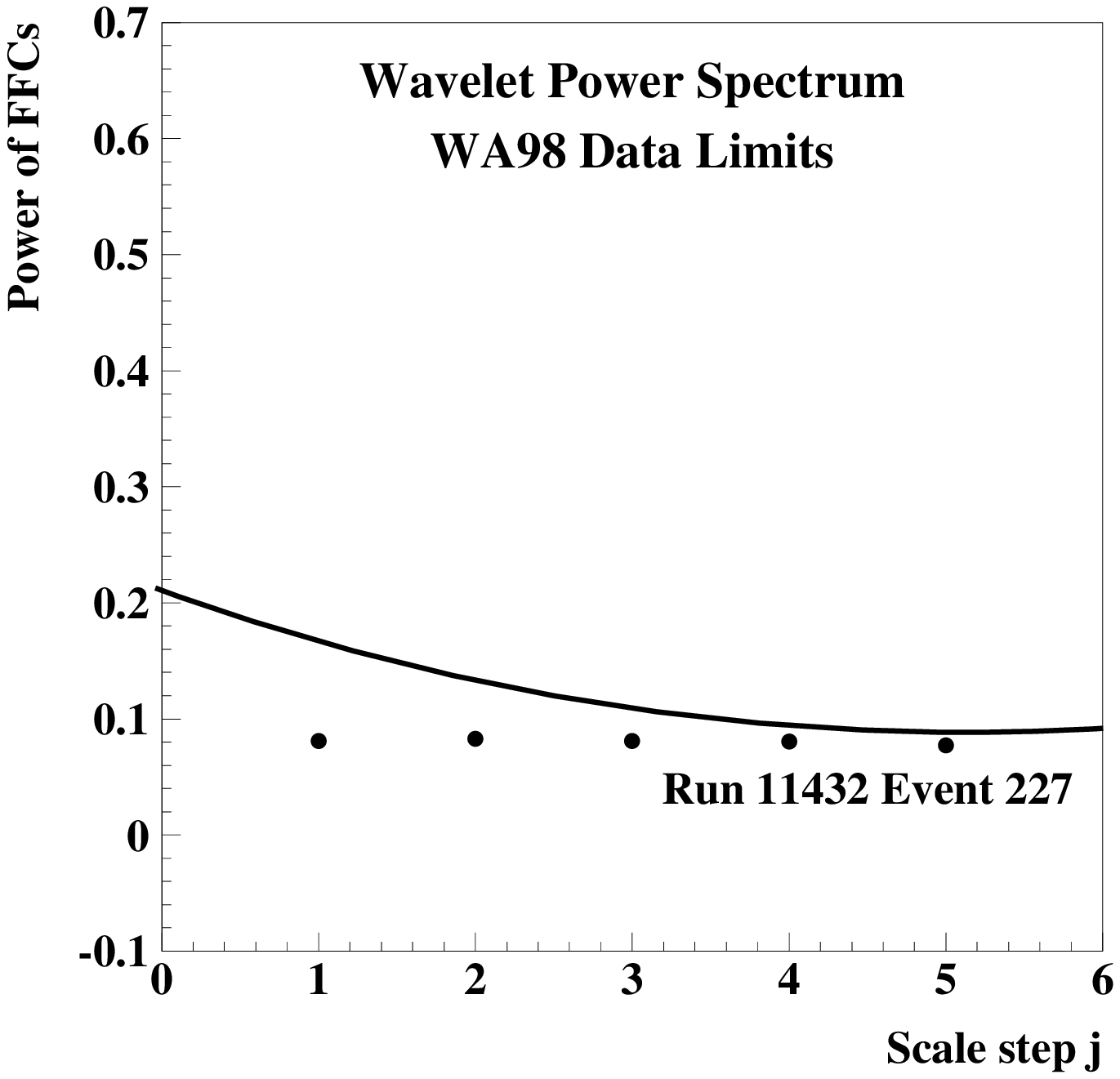}
}
\end{picture}
\vspace{-1.0cm}
\caption{
(a) Wavelet power spectrum from simulated MC events
    for different percentage of DCC events (b) 
    Power spectrum for WA98 data. Events beyond the three
    sigma value of background MC simulation are to be selected as 
    DCC candidates for further study. 
}
\end{figure}

    Another interesting quantity in the fluctuation analysis at each
    scale is the wavelet power spectrum, defined as:
\begin{equation}
       P_j = \frac{1}{2^j} \sum^{2^j-1}_{k=0}\mid \tilde{f}_{jk}\mid ^2.
\end{equation}
    Figure~5(a) shows the power spectrum calculation for
    different cases. The power
    spectrum for background MC simulated events is quite flat at all 
    scales, $j$. In the left hand panel for figure~5 we show simulation
    for 100\% and 10\% DCC events.
    In these cases we are averaging over many events
    which is making the $P_j$ very weak. That means, $P_j$ at 
    j=1 which is so much higher for 100\% DCC events
    has gone down below noticeable level for 10\% DCC. Instead of
    averaging over events which kills
    the signals this plot will be presented event-by-event.

    One can filter out significant number of exotic events by selecting 
    the events which have FFC beyond the three sigma level of the fitted 
    gaussian of the solid curve of figure~5(b).
    The power spectrum corresponding to the 
    three sigma value of FFC is shown in the right panel of the figure as a 
    solid curve. The selected events will have a power value above this
    line. These events, once found, have to be analyzed further by 
    looking at $\eta-\phi$ or $x-y$ event display plots for 
    the existence of DCC. 

\subsection{Bivariate Moments}

Since pions from the DCC domains would appear as localized fluctuations
in the joint charged to neutral particle distributions, one of the
elegant analysis methods is to calculate {\it bi-variate factorial
moments} such as:
\begin{eqnarray}
f_{1,0}(ch, \gamma)  =  \langle N_{ch}\rangle,
\hspace{1cm} f_{0,1}(ch, \gamma)  =  \langle N_{\gamma}\rangle, \nonumber \\
\hspace*{2cm}f_{1,1}(ch, \gamma)  =   \langle N_{ch}N_{\gamma}\rangle,
\hspace{1cm}f_{2,0}(ch, \gamma)  =  \langle N_{ch}(N_{ch}-1)\rangle
\end{eqnarray}
and so on. Here the averaging is done over a large
number of samples. Filtering of true fluctuations from the above
distributions become quite complicated because of various detector
effects, efficiency and background factors. 
Some of the deficiencies could be overcome by using the
generating function technique {\cite {corr}} which can be applied to
concerned distributions.  In that way we can evaluate normalized
factorial moments for the distributions which have the ability to filter
out the non-statistical fluctuations from 
the envelop of statistical noise. One defines the {\it normalized bi-variate
factorial moments} as:
\begin{equation}
F_{i,j} = \frac {f_{i,j}}{f^{i}_{1,0}f^{j}_{0,1}}
=\frac{\left< N_{ch}(N_{ch}-1)\ldots
(N_{ch}-i+1)~N_{\gamma}(N_{\gamma}-1) \ldots (N_{\gamma}-j+1)\right>}
{\left<N_{ch}\right>^{i}\left<N_{\gamma}\right>^{j}}\ ,
\end{equation} 
One defines the ratios of these factorial moments as:
\begin{equation}
                          R_{i,1} = \frac{F_{i,1}}{F_{i+1,0}}
\end{equation}

The Minimax collaboration has shown that these ratios
are not affected by detector effects or efficiencies, and
termed as ``Robust Observables''. Here the sampling is done
over total number of events, rather than event-by-event.
These ratios yield 1 for generic distributions and 1/(i+1) for pure
DCC.

In case of WA98 data the these observables are
not robust, but one can calculate these quantity in case of both
simulation with and without DCC and then compare with data. 
In order to make these quantities suitable for identifying
single event fluctuations the averages are taken over
a considerable number of $\eta-\phi$ phase space bins of
$N_\gamma$ and $N_{\rm ch}$. This can be studied in WA98 because of the
large acceptance of both photon and charged particle detectors.
In average, the results from the data match with MC simulation.
Detailed analysis is in progress to sort out events with large
fluctuations.

\section{Future Prospects}

First we outline some of the new analysis from WA98 experiment
by combining with neutral and charged particle spectra.
Photon and charged particle momentum spectra for different
event classes based on $N_{\gamma-{\rm like}}$ and $N_{\rm ch}$ distributions
(e.g., by choosing events around the mean value of FFC and events
far from the mean) may show significant differences since DCC pions 
are predominantly of low $p_T$. Average $p_T$ of photons obtained 
from PMD may also show differences for these event classes.

Several other signals of DCC have been discussed in the literature
for the detection of DCC.  
It has been proposed that the photon and dilepton signals will be
significantly enhanced in the presence of DCC. This has been discussed
by V.~Koch and by J.~Randrup in this conference.
M.~Asakawa has argued (also in this conference) that non-central 
collisions may be a better place to look for DCC. The two particle
correlation (HBT) effect may also be suppressed if a large domain
forms \cite{gavin2}. Another interesting
signal may be the enhancement in the production of baryons and
antibaryons \cite{kapusta}.

        Most of the future experiments at RHIC and LHC have plans to search
        for the signals of DCC. A highly granular
        photon multiplicity detector is being planned for the STAR 
        experiment at RHIC, which, in combination with
        charged particle detectors and FTPC, will be quite adequate for
        DCC search. Here one would be able to select on the low $p_T$ particles,
        characteristic of DCC pions.
        DCC search in PHENIX will be possible by correlating signals from
        charged particle detectors and photons from EMCAL.
        The PHOBOS experiment, with its large acceptance charged multiplicity
        detector, will be able to search for unusual fluctuations in the 
        production of charged particles. The fluctuations can then be 
        correlated to the production of very low $p_T$ hadrons measured in 
        its two arm spectrometer. The BRAHMS experiment is considering
        to add a photon arm to look for DCC.
        A Photon Multiplicity Detector similar to that of STAR
        is proposed to be included in the ALICE experiment at LHC, which
        along with charged particle detectors will be able to search for DCC.

\section{Summary}

The phenomenon of DCC is quite interesting, its observation in heavy
ion collisions would signal the chiral transition as well as bring a
wealth of information for understanding of QCD. Since the hope is
for such transition to occur at RHIC and LHC energies, we have to be
well equipped
with precision measurement tools and sophisticated analysis
methods. In this article we have outlined a list of analysis methods
that exist at present with a speculation on other signals and new
techniques for future. With continued development and understanding
in theoretical aspects and analysis tools, one would
hope for the best to come from the new data from the upcoming experiments.

\vspace*{0.5cm}
TKN acknowledges very fruitful discussions with Krishna Rajagopal,
Jorgen Randrup, Ajit Srivastava and Xin-Nian Wang.


\begin{thebibliography}{9}

\bibitem{ans1}      A.A. Anselm, M.G. Ryskin, 
                    Phys. Lett. {\bf B266}, (1991), 482.
\bibitem{blai1}     J. -P. Blaizot and A. Krzywcki, 
                    Phys. Rev. {\bf D46} (1992) 246.
\bibitem{bjor1}     J.D. Bjorken, Int. J. Mod. Phys. {\bf A7}, (1992), 4189;
                    J.D. Bjorken, K.L. Kowalski, C.C. Taylor, ``Baked Alaska'',
                    SLAC-PUB-6109, Apr., 1993.
\bibitem{raj1}      K. Rajagopal, F. Wilczek, 
                    Nucl. Phys. {\bf B399}, (1993), 395.
\bibitem{cosmic1}   Proceedings of VIII International Symposium on Very High
                    Energy Cosmic Ray Interactions (Tokyo, Japan), 
                    24-30 July 1994;
                    C.M.G. Lattes, Y. Fujimoto, and S. Hasegawa, 
                    Phys. Rep. {\bf 65}, (1980), 151;
                    J. Lord and J. Iwai, International conference on HEP 
                    (Dallas, 1992)
\bibitem{bjor2}     J.D. Bjorken in hep-ph/9712434.
\bibitem{gavin}     S. Gavin, Andreas Gocksch, Robert D. Pisarski, 
                    Phys. Rev. Lett. {\bf 72} (1994) 2143.
\bibitem{gavin2}    S. Gavin, Nucl. Phys. A590 {\bf 176} (1995) 163.
\bibitem{randrup1}   Jorgen Randrup, Nucl. Phys. {\bf A616} (1997) 531.
\bibitem{randrup2}   Jorgen Randrup and Robert L. Thews, hep-ph/9705260.
\bibitem{asakawa}   Masayuki Asakawa, Zheng Huang, Xin-Nian Wang, 
                    Phys. Rev. Lett. {\bf 74} (1995), 3126. 
\bibitem{raj2}      Krishna Rajagopal,
                    ``The Chiral Phase Transition in QCD: Critical
                    Phenomena and Long Wavelength Pion Oscillations'', 
                    appeared in Quark-Gluon Plasma 2, edited by R. Hwa, 
                    World Scientific, 1995; and Krishna Rajagopal 
                    in hep-ph/9703258.
\bibitem{minimax}   T.C. Brooks et al. (Minimax Collaboration), hep-ph/9609375,
                    J.D. Bjorken et al., hep-ph/9610379.
\bibitem{bolek1}    B. Wyslouch et al., The WA98 Collaboration,
                    these proceedings.
\bibitem{aggar1}    M.M. Aggarwal et al., The WA93 Collaboration,
                    Nucl. Instr. Meth. {\bf A372} (1996) 143.
\bibitem{wa98dcc}   M.M. Aggarwal et al., The WA98 Collaboration,
                    to be published in Phys. Lett. B.
\bibitem{huang}     Zheng Huang et al., Phys. Rev. {\bf D54} (1996) 750.
%, Ina Sarcevic, Robert Thews and 
%Xin-Nian Wang, Phys. Rev. {\bf D54} (1996) 750.
%\bibitem{wer1}      K. Werner, Phys. Lett. {\bf B208}, (1988), 520.
\bibitem{corr}      E. A. DeWolf, I. M. Dremin, W. Kittel, Phys. Rep. 270: 1
                    (1996).
\bibitem{kapusta}   J.I. Kapusta and A.M. Srivastava, 
                    Phys. Rev. D {\bf 52} (1995) 2977.
\end{thebibliography}
\end{document}